\documentclass[useAMS]{mnras}

\usepackage{graphicx}
\usepackage[fleqn]{amsmath}
\usepackage{multirow}
\usepackage{hyperref}
\usepackage{xspace}
\usepackage{url}
\usepackage{color}

    \newcommand{\msun}{$M_{\hbox{$\odot$}}$\xspace}

    \renewcommand{\arcsec}{\hbox{$^{\prime\prime}$}\xspace}
    \renewcommand{\degr}{\hbox{$^\circ$}\xspace}
    \newcommand{\kms}{\hbox{km s$^{-1}$}\xspace}
    \newcommand{\fluxcgs}{\hbox{erg s$^{-1}$ cm$^{-2}$}\xspace}

    \newcommand{\lumcgs}{\hbox{erg s$^{-1}$}\xspace}
    \newcommand{\ang}{\AA\xspace}
    \newcommand{\fek}{Fe\,K\xspace}

    \newcommand{\ka}{K$\alpha$\xspace}

    \newcommand{\hb}{H$\beta$\xspace}
    
    \newcommand{\feii}{Fe\,\textsc{ii}\xspace}
    \newcommand{\fexxv}{Fe\,\textsc{xxv}\xspace}
    \newcommand{\fexxvi}{Fe\,\textsc{xxvi}\xspace}
    \newcommand{\civ}{C\,\textsc{iv}\xspace}
    \newcommand{\oiii}{[O\,\textsc{iii}]\xspace}
    
    \newcommand{\vout}{v_{\rm out}}

    \newcommand{\dchi}{\Delta\chi^{2}}
    
    \newcommand{\chidof}{\chi^{2}/\nu}
    \newcommand{\redchi}{\chi^{2}_\nu}
    \newcommand{\nh}{N_{\rm H}}

    \newcommand{\lbol}{L_{\rm bol}}

    \newcommand{\mbh}{M_{\rm BH}}

    \renewcommand{\rmn}[1]{{\mathrm{#1}}}
    \newcommand{\xmm}{\emph{XMM--Newton}\xspace}

    \newcommand{\chandra}{\emph{Chandra}\xspace}

    \newcommand{\rosat}{\emph{ROSAT}\xspace}
    \newcommand{\optxagnf}{\texttt{optxagnf}\xspace}
    
    \newcommand{\xstar}{\textsc{xstar}\xspace}
    \newcommand{\xspec}{\textsc{xspec}\xspace}

    \newcommand{\ftools}{\textsc{ftools}\xspace}
    \newcommand{\sas}{\textsc{sas}\xspace}

    \newcommand{\rmfgen}{\texttt{rmfgen}\xspace}
    \newcommand{\arfgen}{\texttt{arfgen}\xspace}
    \newcommand{\fakeit}{\texttt{fakeit}\xspace}

    \newcommand{\ton}{Ton\,28\xspace}

\title[The X-ray wind of the quasar Ton\,28]{Towards an informed quest for accretion disc winds in quasars: the intriguing case of Ton\,28}
\author[Nardini, Lusso \& Bisogni]
{E. Nardini,$^1$\thanks{E-mail: enardini@arcetri.astro.it} E. Lusso$^{2}$ and S. Bisogni$^{3,1}$\\ 
$^1$INAF -- Osservatorio Astrofisico di Arcetri, Largo Enrico Fermi 5, I-50125 Firenze, Italy\\
$^2$Centre for Extragalactic Astronomy, Department of Physics, Durham University, South Road, Durham DH1 3LE, UK\\
$^3$Harvard-Smithsonian Center for Astrophysics, 60 Garden Street, Cambridge, MA 02138, USA}

\begin{document}

\date{Released Xxxx Xxxxx XX}

\pagerange{\pageref{firstpage}--\pageref{lastpage}} \pubyear{2018}

\maketitle

\label{firstpage}

\begin{abstract}
We report on the detection of a blueshifted \fek absorption feature in two consecutive \xmm observations of the luminous blue quasar \ton, at the 4$\sigma$ cumulative significance. The rest energy of 9.2 keV implies the presence of an accretion disc wind with bulk outflow velocity of $\sim$0.28$c$, while the kinetic power is most likely a few per cent of the quasar luminosity. Remarkably, \ton had been specifically selected as an optimal target to reveal an ultra-fast X-ray wind based on its total luminosity ($\lbol > 10^{46}$ \lumcgs) and \oiii$\lambda$5007\,\ang equivalent width ($\rmn{EW} < 6$ \ang), suggestive of high accretion rate and low inclination, respectively. Other peculiar optical/UV emission-line properties include narrow \hb, strong \feii and blueshifted \civ. These are key parameters in the Eigenvector 1 formalism, and are frequently found in active galaxies with ongoing accretion disc winds, hinting at a common physical explanation. Provided that the effectiveness of our selection method is confirmed with similar sources, this result could represent the first step towards the characterization of black-hole winds through multiwavelength indicators in the absence of high-quality X-ray spectra.
\end{abstract}


\section{Introduction}

The assembly of a supermassive black hole (BH) in the core of a galaxy is expected to heavily affect the properties and evolution of the entire system, if just a minor fraction of the energy released as a by-product of the accretion process can efficiently couple with the gas in the host (e.g. Silk \& Rees 1998; King 2003; Fabian 2012). Galaxy-wide outflows are now commonly found in the most luminous active galactic nuclei (AGN), and their power source must be ultimately linked to the BH growth itself. At high luminosity, radiation pressure will naturally strip matter off the accretion flow (e.g. Proga 2007). The ensuing ultra-fast winds can be revealed through highly ionized iron absorption features in the X-ray spectra, and are widely regarded as the onset of AGN feedback (Nardini et al. 2015, and references therein). Very little is known, however, on how the BH wind power is transferred from sub-pc to kpc scales in practice. 

A major impediment to any coherent picture is that so far most detections of X-ray winds have been achieved through either blind searches (Tombesi et al. 2010; Gofford et al. 2013) or follow-ups of peculiar objects (Chartas et al. 2002; Giustini et al. 2011; Lanzuisi et al. 2012), maintaining an undertone of \textit{fortuity}.
It is thus desirable to understand whether ongoing X-ray winds leave any distinctive signatures imprinted at other wavelengths on the intrinsic AGN spectrum. Such a prospect is particularly compelling for the study of high-redshift sources accreting around the Eddington limit, for which high-quality X-ray spectra (and hence access to the usual wind tracers) will not be available until the advent of the next-generation observatories. The results presented in this Letter stem from the effort to single out a quasar in its `X-ray blow-out' evolutionary phase. 

\section{Target selection and Observation}

Our search for an optimal target started from the necessity to have a clean line of sight to the nuclear regions, avoiding reddening and obscuration effects that would complicate continuum and line measurements. We therefore considered the 1091 quasars with equivalent width (EW) of the \oiii$\lambda$5007\ang line smaller than 6 \ang from the sample analysed by Bisogni, Marconi \& Risaliti (2017), extracted from the Sloan Digital Sky Survey (SDSS) 7th Data Release. The \oiii EW is a reliable indicator of the accretion disc orientation, with the lowest values corresponding to nearly face-on configurations. Next, we restricted our attention to the subset of objects with $\lbol > 10^{46}$ \lumcgs, as undisputable and persistent ultra-fast X-ray winds are mostly found at the higher end of the AGN luminosity function at any redshift (e.g. Chartas et al. 2009; Matzeu et al. 2017), whereas \fek absorption features in local Seyferts have marginal significance, lower blueshifts and/or transient nature (Ponti et al. 2013; Braito et al. 2014; Marinucci et al. 2018). The latter cut leaves 145 sources. We further imposed a \rosat detection, whose merit is twofold: providing information on the X-ray brightness, and minimizing the chance of X-ray obscuration. Of the 43 remaining sources, 36 are radio-quiet: we chose the one with the highest X-ray count rate, \ton.

\begin{figure}
\includegraphics[width=8.5cm]{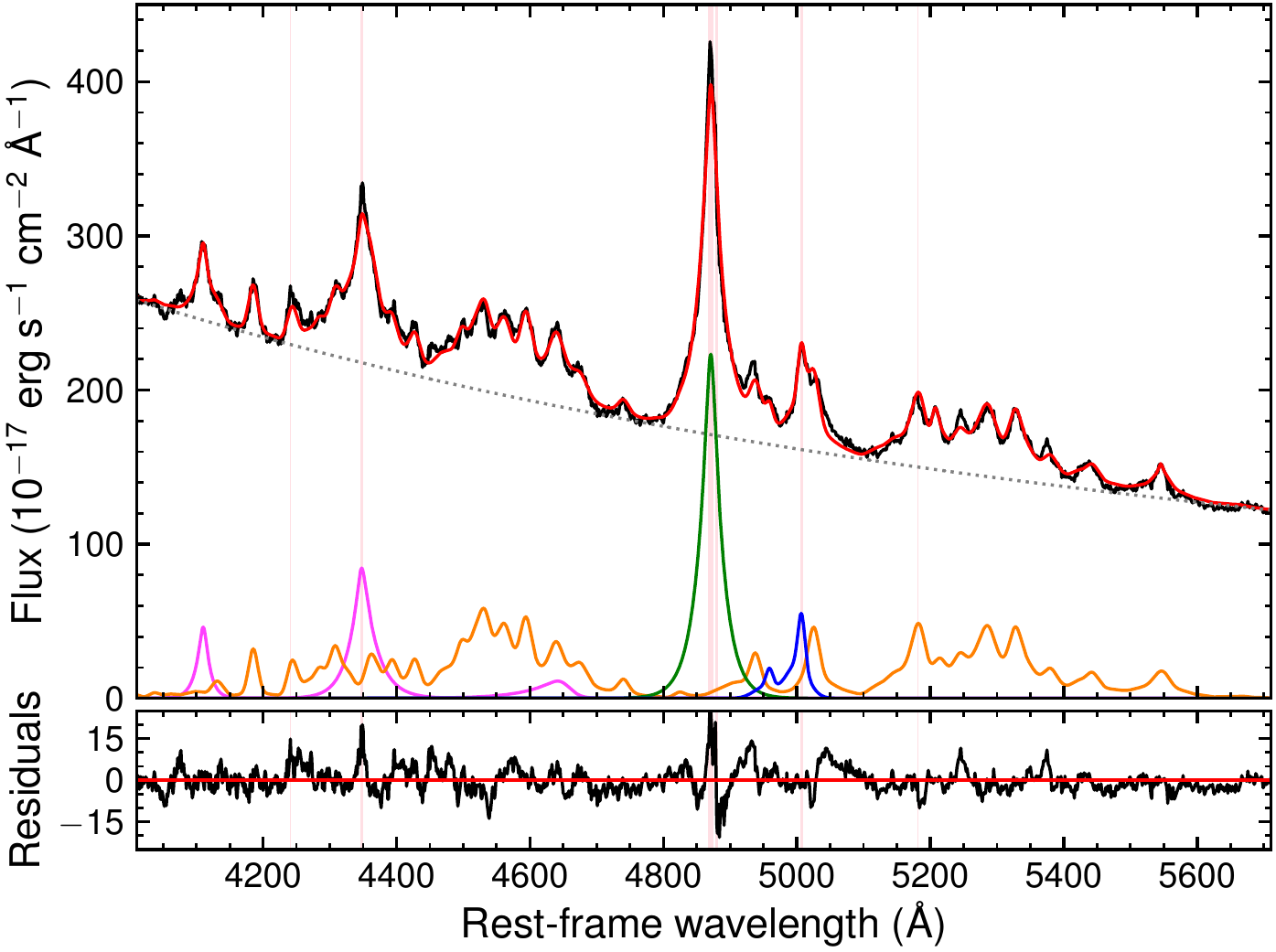}
\vspace*{-10pt}
\caption{Zoom in onto the \hb region of the SDSS spectrum of \ton. Best-fit model (red curve) and residuals (bottom panel) are also shown. The main emission components are plotted with different colours: \hb in green, total \oiii$\lambda$$\lambda$4959,\,5007\,\ang (narrow plus broad, blueshifted) in blue, total \feii in orange, other lines (H$\delta$, H$\gamma$, He\,\textsc{ii}\,$\lambda$4686\,\ang) in magenta. The shaded bands correspond to masked pixels. See Bisogni et al. (2017) for more details.}
\label{os}
\end{figure}

\ton is among the most luminous ($\lbol \sim 2.5 \times 10^{46}$ \lumcgs; Shen et al. 2011) non-jetted quasars at relatively low redshift ($z \simeq 0.329$), whose optical and ultraviolet (UV) spectra have been extensively examined in the literature, mainly for its nature of background beacon (e.g. Danforth et al. 2016). A portion of its SDSS spectrum is shown in Fig.~\ref{os}. The \feii emission strength is a direct consequence of our selection, given its well-known anti-correlation with \oiii (Boroson \& Green 1992). In the UV, the \textit{Hubble Space Telescope} spectrum exhibits an asymmetric \civ$\lambda$1549\,\ang profile, blueshifted by 1120 \kms at the full-width at half maximum (FWHM) centroid, which was kinematically modelled alongside the blueshifted \oiii component (680 \kms) as arising from the same radial outflow of conical shape, with half-opening angle of 85\degr and inclination of 15\degr (Zamanov et al. 2002). In the empirical framework designed to organize and explain the diversity of quasars known as `Eigenvector 1' (see Sulentic et al. 2000, 2007 for details), all these optical/UV spectral properties point at the specific region of the main sequence occupied by extreme population A sources (Marziani et al. 2003), suggesting a possible route to further refine our selection criteria. By virtue of its luminosity, \ton also stands out as a potential \textit{twin} to PDS\,456, the prototype of quasar accretion disc wind (Simpson et al. 1999; Reeves, O'Brien \& Ward 2003; O'Brien et al. 2005). Despite this, it had been completely overlooked in the X-rays. 

In order to obtain the first broad-band X-ray view of \ton, we were awarded an 85 ks \xmm observation in cycle 16 (PI: E. Lusso), representing the longest continuous exposure of the source that could be achieved at that time. The observation was actually split in two blocks of similar duration and performed over consecutive orbits, on 2017 May 14--16. The EPIC instruments were operated in Large Window mode with medium optical filter, and the corresponding event files were reprocessed within the Science Analysis System (\sas) v16.1.0. In this Letter we only concentrate on the EPIC/pn spectra, as the effective area of the two MOS detectors quickly falls off at high energy and does not provide enough statistics for our analysis. The source spectra were extracted from circular regions of 30\arcsec radius around the position of the target, while the background was estimated over a nearby 60\arcsec circle situated on the same chip. Single and double pixel events (patterns 0--4) were selected. Redistribution matrices and ancillary response files were generated, respectively, with \rmfgen and \arfgen.

The first observation (Obs.\,1) was significantly affected by background flares. The standard filtering criterion, based on the fiducial acceptable threshold for the overall 10--12 keV count rate, turns out to be overly conservative in this case, entailing the rejection of about 41 per cent of the net exposure time. We have therefore followed an optimized procedure aimed at maximizing the spectral quality in the 2--8 keV band. Specifically, only the periods that cause the degradation of the signal-to-noise ratio in the energy range of interest were discarded (see Piconcelli et al. 2004 for a more detailed description of this method). This allowed us to recover a further 25 per cent of the total exposure compared to the standard cut, for a good time interval of 31.5 ks. Apart from a very minor flare, the background was low and stable for the entire span of the second observation (Obs.\,2), so the full exposure of 41.4 ks is available (Table~\ref{t1}).

\begin{table}
\centering
\small
\caption{Log of the \xmm observations of \ton.}
\label{t1}
\begin{tabular}{l@{\hspace{10pt}}c@{\hspace{10pt}}c@{\hspace{10pt}}c}
\hline
Obs.\,ID & Date\,--\,Time\,(UTC)$^a$ & Exp.\,(ks)$^b$ & Counts$^c$ \\
\hline
0804560101 & 2017\,May\,14\,--\,06:10:55 & 31.462 & 28392 \\
0804560201 & 2017\,May\,15\,--\,23:37:04 & 41.369 & 35979 \\
\hline
\end{tabular}
\flushleft
\textit{Notes.} $^a$Exposure start. $^b$Net exposure. $^c$Net 0.3--10 keV counts.
\end{table}

The spectral analysis has been performed with the \xspec v12.9.1 fitting package. The data were binned to ensure a significance of at least 4$\sigma$ per energy channel, and the uncertainties are given at the 90 per cent confidence level ($\dchi = 2.71$) for the single parameter of interest, unless stated otherwise. Since the spectral variability is negligible (see below), we also created a merged spectrum and the relative response files with the usual \ftools tasks, which have been used for a consistency check of our results. The photometric fluxes from the five requested Optical Monitor (OM) filters (all except UVW2) have also been retrieved for comparison with the SDSS spectrum and the X-ray emission.

\section{Data Analysis}

\begin{figure}
\includegraphics[width=8.5cm]{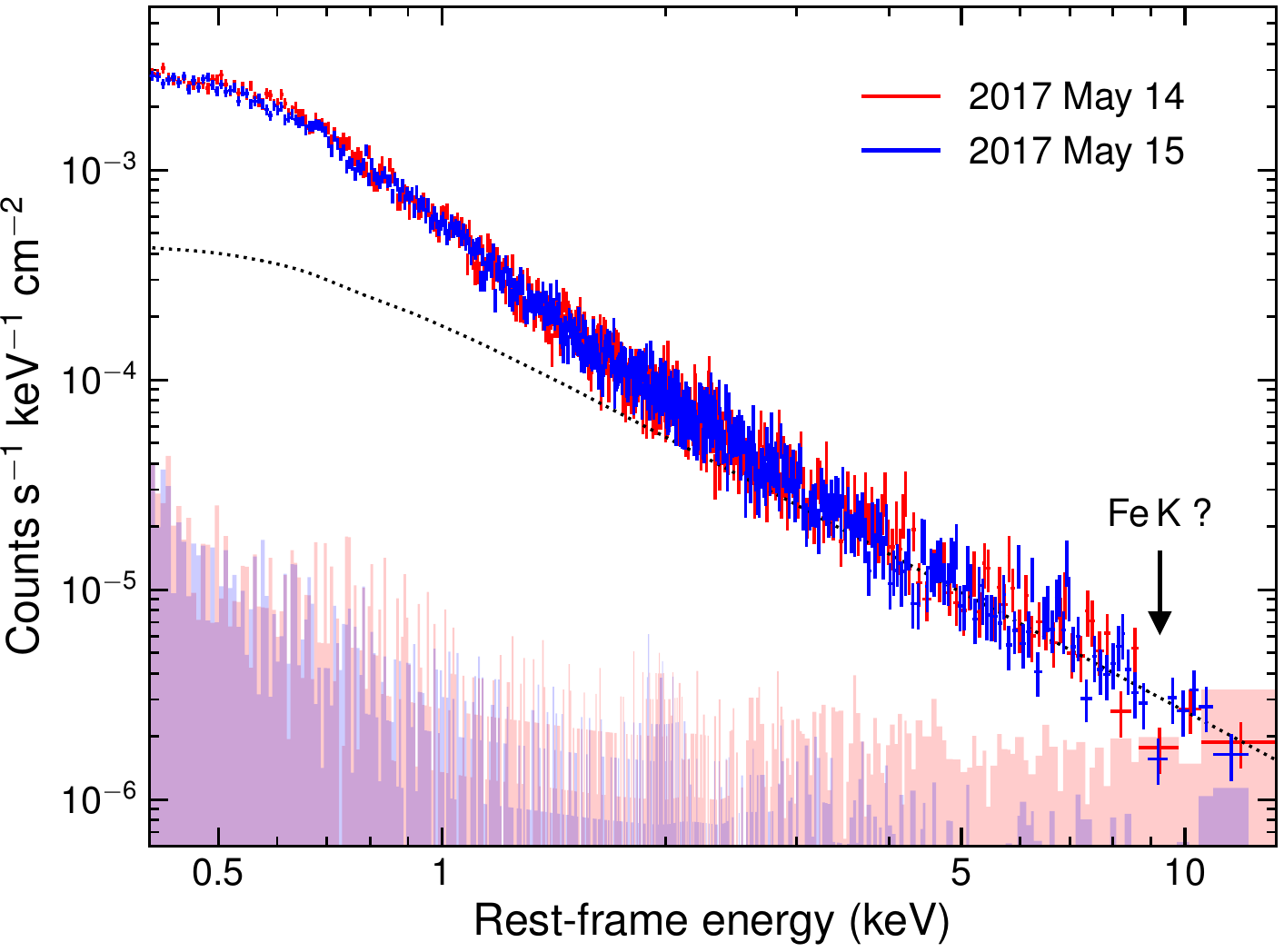}
\vspace*{-10pt}
\caption{EPIC/pn spectra of \ton from the first (red) and the second (blue) \xmm observation. A power law with $\Gamma = 1.9$, only modified by the Galactic column density at low energies, is also plotted for reference (black dotted curve). This brings out the prominent soft excess and a possible absorption feature at about 9.2 keV, as indicated by the arrow. Note that the latter feature appears to be present in both spectra, despite the markedly different background level (shaded regions) in the hard X-rays between the two observations.}
\label{xs}
\end{figure}

The EPIC/pn spectra from the two observations of \ton are shown in Fig.~\ref{xs}. The simple visual inspection reveals that the continuum has the typical traits of an X-ray unobscured quasar, where the prominent soft excess is more intense than the extrapolation of the hard X-ray power-law tail by about a factor of six at 0.5 keV. For a preliminary, yet physically motivated characterization of the broad-band X-ray emission we therefore made use of the \optxagnf model (Done et al. 2012), which self-consistently combines warm and hot Comptonization from the inner accretion disc atmosphere and the X-ray corona, respectively originating the soft excess and the hard power law, and also accounts for thermal emission from the outer disc. In particular, with \optxagnf we can probe the shape of the ionizing continuum, whose convenience will become clearer later on. A more detailed analysis, integrated in the multiwavelength perspective, will be performed in a forthcoming companion paper. 

A reliable BH mass estimate is highly desirable for a direct determination of the accretion rate, which is a key parameter of \optxagnf. A measure of $\mbh \sim 3 \times 10^7$ \msun has been recently reported by Du et al. (2018), based on a reverberation mapping campaign in which \ton was monitored twice one year apart. An \hb lag, however, was seen only once, and its uncertainty is anomalously large. We then tried to derive an alternative $\mbh$ value from the single-epoch broad-line region radius versus luminosity relations of Bentz et al. (2013) and our decomposition of the optical spectrum (Fig.~\ref{os}). With FWHM\,(\hb) $= 1877$\,($\pm$131) \kms and $\lambda L_\lambda (5100$\,\ang) $\simeq 2.8 \times 10^{45}$ \lumcgs, we get $\mbh \sim 1.3 \times 10^8$ \msun, for which accretion would still occur at 1.5 times above the Eddington limit. This larger BH mass improves the connection between the OM data points, which are found to be in very good agreement with the SDSS spectrum, and the X-ray emission in a joint fit. Yet, the power law is employed to reduce the relative strength of the soft excess, and its slope is clearly too steep at 3--10 keV (where $\redchi \sim 1.3$). This might hint at the presence of some reflected continuum, but spectral coverage above 10 keV would be needed to constrain its strength. For our current purpose of evaluating the optical to hard X-ray spectral energy distribution (SED) of \ton, $\mbh$ was left free to vary in \optxagnf.

\begin{figure}
\includegraphics[width=8.5cm]{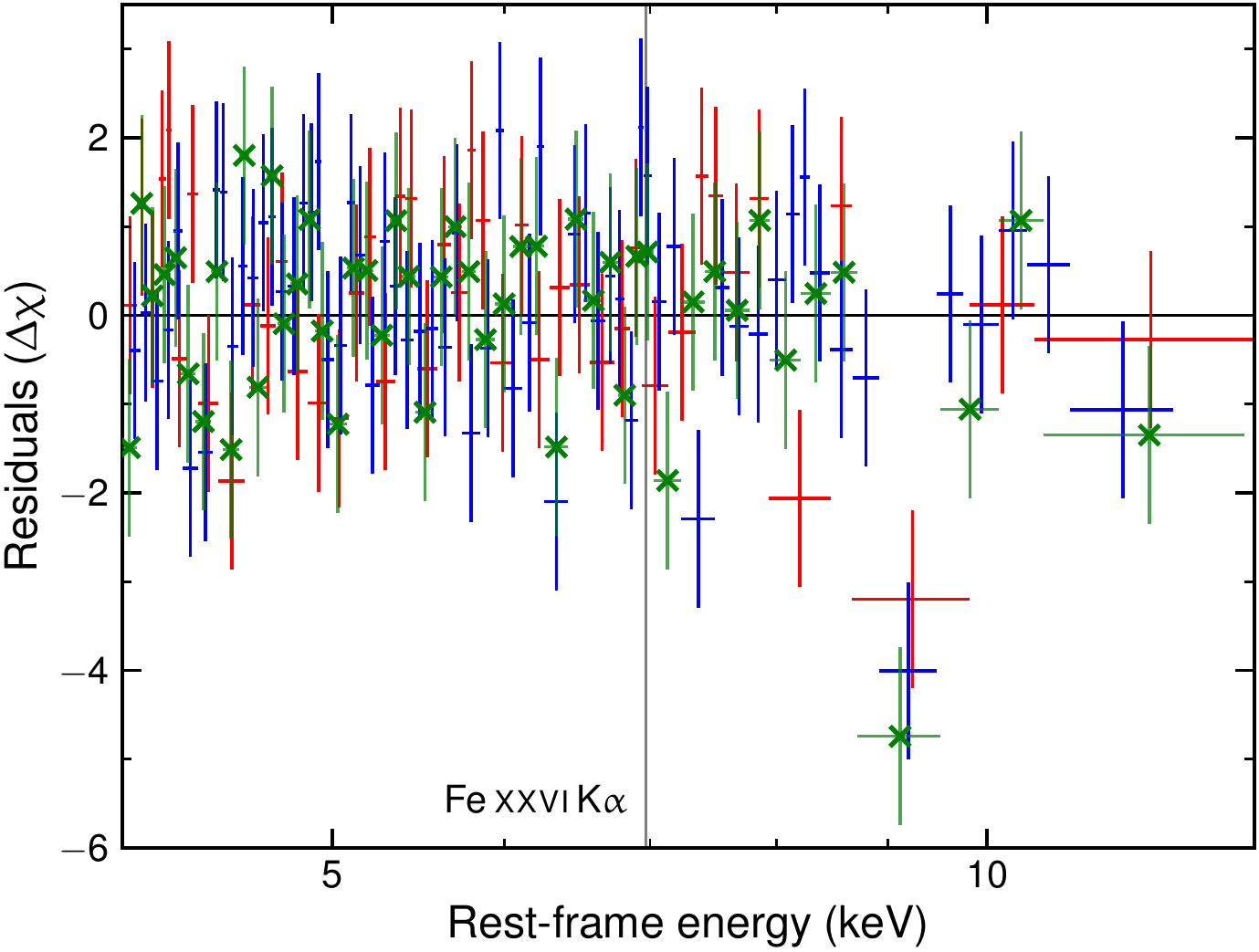}
\vspace*{-10pt}
\caption{Residuals in units of standard deviations ($\sigma$) after removing the blueshifted absorption feature from the best-fitting model at 3--10 keV (observed). Colours are as follows: red (Obs.\,1), blue (Obs.\,2), green (merged). The vertical line indicates the expected position of the \fexxvi \ka transition at rest. The latter identification implies an outflow velocity of $\vout \sim 0.25$--0.30$c$.}
\label{wf}
\end{figure}

Irrespective of the broad-band interpretation, negative residuals remain in both spectra at about 6.9 keV (9.2 keV in the rest-frame; Fig.~\ref{wf}). To better examine the properties of this possible absorption feature, we focussed on the 3--10 keV band, following the common practice of modelling the local continuum only (e.g. Tombesi et al. 2010).  The spectra from Obs.\,1 and Obs.\,2 were fitted separately, yielding remarkably consistent results in spite of their different quality. The continuum and line parameters are listed in Table~\ref{t2}. For a provisional identification with the \fexxvi \ka at 6.97 keV, the implied outflow velocity is of the order of $\vout \sim 0.25$--0.30$c$. When all the parameters are tied between the two observations, the improvement in the joint fit statistics is $\dchi = -19.5$ for the loss of two degrees of freedom. According to an $F$-test, the line detection is genuine with a probability of 99.97 per cent, or 3.7$\sigma$.

The lack of any obvious \ka emission feature from neutral iron at 6.4 keV can be explained in the context of the X-ray Baldwin effect (Iwasawa \& Taniguchi 1993). However, any model accepts a narrow line with rest EW of $\sim$140\,($\pm$80) eV, centred at an energy that does not formally correspond to any major transition, $E = 6.86^{+0.09}_{-0.11}$ keV. We tentatively rule out any residual calibration inaccuracy after the Charge Transfer Inefficiency correction, as this would also affect the range around the gold M edge, and much more severely (see e.g. Nardini et al. 2016). Unless moderate velocity shifts are involved, the line could then be a superposition of fluorescent \ka lines from Fe\,\textsc{xxv--xxvi}, confirming the existence of highly ionized gas in the nuclear regions of \ton. Interestingly, also the absorption feature at 9.2 keV is most likely a blend of the same transitions (see below). While this leaves room for speculation (i.e. reflection off the wind), we do not discuss the emission line any further in this work. 

\begin{table}
\centering
\small
\caption{Best-fitting absorption line parameters when modelled with a Gaussian profile against the local power-law continuum. $F$ is the flux over the observed 3--10 keV band, while $\dchi$ is the statistical improvement after the inclusion of the line (for the loss of two degrees of freedom). The line's rest energy is frozen (f), and its rest EW is $-$340\,($\pm 160$) eV (tied, 90 per cent confidence).}
\label{t2}
\begin{tabular}{l@{\hspace{16pt}}c@{\hspace{8pt}}c@{\hspace{8pt}}c}
\hline
Parameter & Obs.\,1 & Obs.\,2 & Tied \\
\hline
$E_\rmn{abs}$\,(keV) & 6.97(f) & 6.97(f) & 6.97(f) \\[0.5ex]
$\vout/c$ & $0.29^{+0.03}_{-0.06}$ & $0.25^{+0.04}_{-0.01}$ & $0.25^{+0.04}_{-0.01}$ \\[0.5ex]
$A_\rmn{abs}$\,(10$^{-6}$ s$^{-1}$ cm$^{-2}$) & $-1.7^{+1.0}_{-1.3}$ & $-1.0^{+0.5}_{-0.6}$ & $-1.2^{+0.5}_{-0.4}$ \\[0.5ex]
$\Gamma$ & $1.95^{+0.25}_{-0.25}$ & $1.81^{+0.20}_{-0.19}$ & $1.88^{+0.15}_{-0.16}$ \\[0.5ex]
$F$\,(10$^{-13}$ \fluxcgs) & $3.7^{+0.3}_{-1.0}$ & $3.6^{+0.2}_{-0.4}$ & $3.7^{+0.1}_{-0.3}$ \\[0.5ex]
$\chidof$ & 44.0/38 & 63.3/58 & 111.6/102 \\[0.5ex]
$\dchi$ & $-7.6$ & $-12.9$ & $-19.5$ \\[0.5ex]
$F$-test probability & 0.9514 & 0.9954 & 0.9997 \\
\hline
\end{tabular}
\end{table}

In the case of outflows, where a line's energy is clearly not known in advance, a mere $F$-test is not appropriate to establish the reliability of a feature, as spurious detections can emerge at any energy (e.g. Protassov et al. 2002). Hence, in order to corroborate our findings, we also resorted to Monte Carlo simulations. For simplicity, at this stage we made use of the merged spectrum, grouped to a significance of 6$\sigma$ per energy bin. Here the addition of a Gaussian absorption profile to the power-law continuum brings an improvement of $\dchi = -17.3$ for the loss of two degrees of freedom\footnote{We note that this value is not sensitive to the spectral grouping. If 5$\sigma$ bins were adopted instead, we would still get $\dchi = -17.0$.} (see Fig.~\ref{wf}). We neglected the emission line and assumed as null hypothesis a featureless power law, with photon index and normalization as derived from the best fit over the 3--10 keV band. Following Miniutti \& Fabian (2006), we ran a preliminary simulation with the \fakeit command within \xspec to generate a spectrum with the same exposure (72.8 ks), background and response of the real data. We performed a first fit to obtain a refined input model, taking into account the effects of photon statistics on the null hypothesis itself, and proceeded with the proper simulation. After applying a self-consistent 6$\sigma$ binning, the resulting spectrum was fitted with a power law, and the reference $\chi^2$ value was recorded. We subsequently scanned the 7--10 keV rest-frame energy band in steps of 0.1 keV for the presence of a line, allowing for both negative and positive amplitudes. The minimum $\chi^2$ returned by the line scan was stored for comparison with the reference value. The whole procedure was repeated for 10,000 times. In only nine cases (eight in absorption, one in emission) the statistical improvement afforded by a spurious line is $\dchi < -17.3$, thus setting the confidence level of our detection to 99.91 per cent, or 3.3$\sigma$. This is still a rather conservative estimate, as the coincidence of residuals at the same energy in two different spectra, although individually less significant (Fig.~\ref{wf}), is not considered; moreover, the false positives in the simulations are all confined below 8.5 keV, boosting the reliability of a feature at 9.2 keV even further.

\section{Discussion and Conclusions}

\begin{figure}
\includegraphics[width=8.5cm]{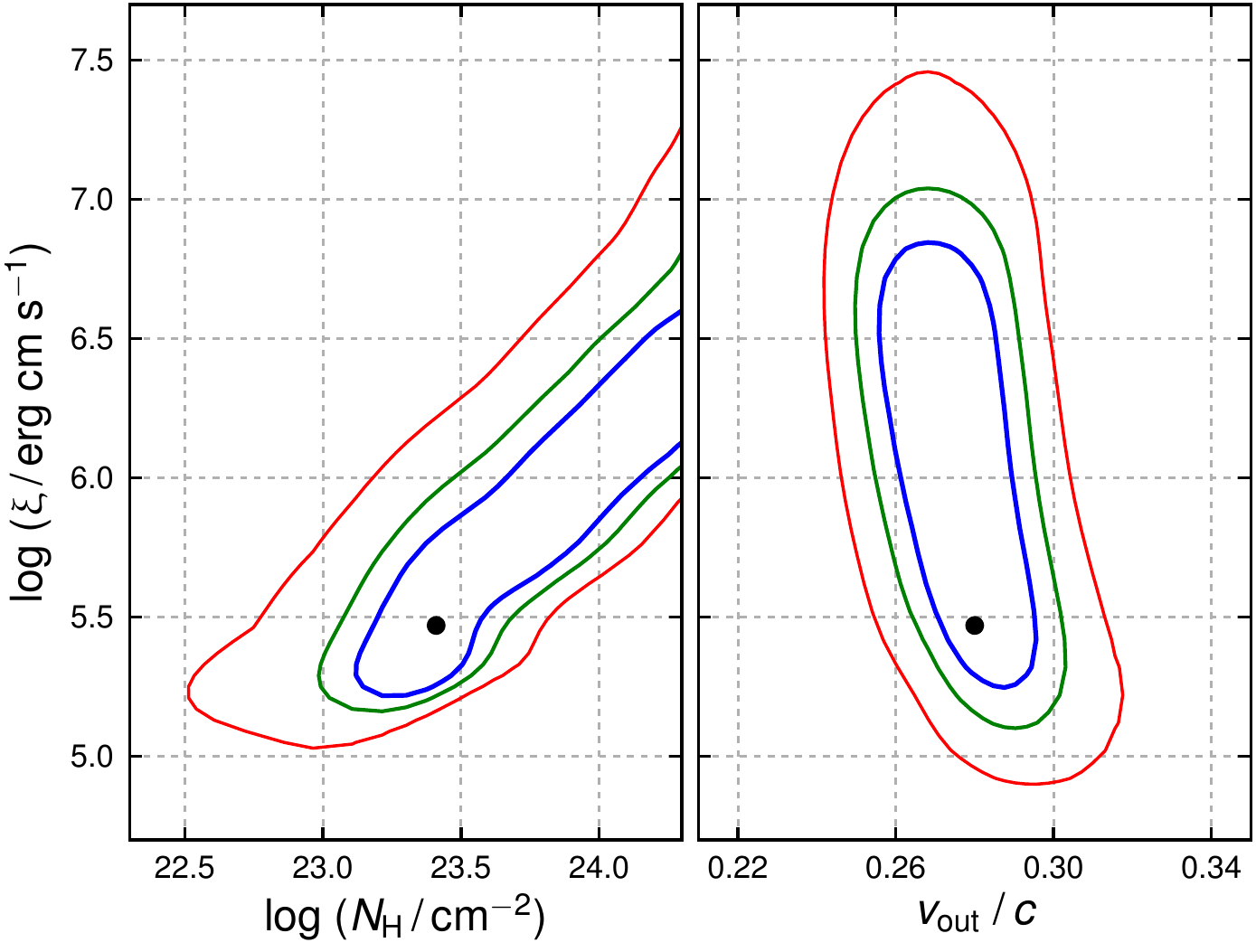}
\vspace*{-10pt}
\caption{Confidence contours at the 68, 90 and 99 per cent level of the ionization parameter versus the column density (left) and the outflow velocity (right) of the absorbing gas. While there is a strong degeneracy in the $\xi$--$\nh$ plane, the outflow velocity is constrained relatively well, as solutions with a dominant contribution from \fexxvi to the 9.2-keV feature are clearly preferred.}
\label{cc}
\end{figure}

Based on our past experience and extensive simulations, a safe (i.e. $>3\sigma$) detection of an unresolved \fek wind signature with $\vout \sim 0.1c$ and $\rmn{EW} = -100$ eV in the EPIC/pn spectrum of a local AGN requires a 2--10 keV flux approaching 10$^{-12}$ \fluxcgs and at least 2,000 net counts at 5--10 keV, for a power-law continuum of $\Gamma = 2$. The exceptional strength of its soft excess (whose value, corrected for Galactic absorption, exceeds 1.2 adopting the definition of Vasudevan et al. 2013) implies that the hard X-ray flux of \ton is considerably lower than expected from the \rosat count rate, falling below the above reference threshold by a factor of two (cf. Table~\ref{t2}). The same applies to the number of 5--10 keV counts available, which is just above 800 from Obs.\,1 and Obs.\,2 combined. None the less, the sizeable rest EW of $-$340\,($\pm 160$) eV and the cosmological redshift of the source (which virtually offsets the more substantial blueshift of the line) proved to be instrumental in stretching the standard limits of detectability. Indeed, the evidence for a highly ionized, mildly relativistic BH wind in \ton appears to be rather robust, as it does not simply rely on the sheer statistical improvement in the spectral fits after the inclusion of the absorption line, but it is also supported independently by Monte Carlo simulations and by the presence of the feature in two observations with strikingly different background intensities, when any other line-like residual due to photon noise is cancelled out in the merged spectrum.

We can therefore attempt to gain more insights into the physical properties of this X-ray wind. To this aim, it is necessary to properly account for absorption by photo-ionized gas, as the outflow velocity itself depends on the identification of the ionic species and transition involved. We first reverted to the full 0.3--10 keV spectra, using a phenomenological continuum model of the form \texttt{zbbody\,+\,bknpower} to overcome the limitations of \optxagnf and accurately reproduce the hard X-ray slope. The marginal differences between Obs.\,1 and Obs.\,2 are ascribed to a $\sim$10 per cent change in the normalization of the blackbody component, while all the other parameters were tied. The fit is indeed very good, with $\chidof = 788.8/823$ before allowing for the wind. Incidentally, the blackbody has a temperature $kT = 116$\,($\pm 3$) eV, while the power law breaks from $\Gamma \simeq 2.5$ to 1.8 around 4.3 keV (rest-frame). We then created a suitable absorption grid with \xstar (Kallman \& Bautista 2001), adopting as ionizing continuum the broad-band SED shape derived from the \optxagnf model with free $\mbh$, coarsely sampled by 10 logarithmically spaced points between 1 eV and 100 keV and normalized to a 1--1000 Rydberg luminosity of 10$^{46}$ \lumcgs. Column density and ionization parameter (defined as $\xi = L/nR^2$, in units of erg cm s$^{-1}$) turn out to be largely degenerate (Fig.~\ref{cc}, left panel). Even so, for solar iron abundance the best nominal $\nh$ hits the hard limit imposed in our grid (2\,$\times$\,10$^{24}$ cm$^{-2}$). We thus opted for moderate iron overabundance ($Z_\rmn{Fe} = 3$), which might even be more appropriate for the innermost regions of an object like \ton (e.g. Hamann \& Ferland 1999). The fits are poorly sensitive to turbulent velocity, which was fixed to 5,000 \kms. The application of such a grid delivers a final $\chidof = 768.9/820$ ($\dchi = -19.9$), confirming the high statistical significance of the accretion disc wind.\footnote{Given the $\xi$--$\nh$ degeneracy, only two degrees of freedom are effectively lost. The improvement corresponding to an $\nh$ fixed to 5\,$\times$\,10$^{23}$ cm$^{-2}$, for instance, is $\dchi = -19.7$, equivalent to 4$\sigma$.} 

We note that this exercise does not yet provide sufficiently tight constraints on the absorber to determine the energetics of the wind, as the mass outflow rate depends on quantities that are either highly uncertain (column density, iron abundance) or completely unknown (covering factor, radial location). Only the outflow velocity is pinpointed with fairly good precision to $\vout = 0.28^{+0.02}_{-0.03}\,c$ (Fig.~\ref{cc}, right panel). \fexxvi is in fact the dominant species over most of the relevant ionization range. For the best-fitting value of $\log\,\xi \simeq 5.5$, the 9.2-keV feature is actually a blend of the \ka lines from \fexxvi and \fexxv, with an approximate intensity ratio of 7\,$:$\,5. For reference, the relative weight is unity at $\log\,\xi \sim 5.2$, while traces of \fexxv are found up to $\log\,\xi \sim 7$. Capitalizing on the measure of the outflow velocity, we can still infer a conservative figure of the kinetic power of the wind, $\dot{\mathcal{E}}_\rmn{kin}$. We use the relations in Nardini \& Zubovas (2018), which assume the escape radius as launch site and a covering factor of 0.5, so that $\dot{\mathcal{E}}_\rmn{kin} \propto \nh \mbh \vout$. For our best guess of $\mbh \sim 10^8$ \msun and $\nh \sim 2 \times 10^{23}$ cm$^{-2}$, we obtain $\dot{\mathcal{E}}_\rmn{kin} \sim 2 \times 10^{44}$ \lumcgs, i.e. about one per cent of the quasar bolometric luminosity. Despite the large uncertainties, it seems then likely that the ultra-fast X-ray wind in \ton meets the minimum energetics requirements for AGN feedback to work (e.g. Hopkins \& Elvis 2010).

While blind searches and occasional detections have already suggested that the appearance of blueshifted \fek absorption is a prevalent phenomenon among AGN, the case of \ton is brought to the next level by the very way our target had been selected. In a sense, this result could be regarded as the first step towards an informed discovery of ultra-fast X-ray winds, and the indirect identification of objects that are currently undergoing such a phase. Our selection was indeed quite straightforward, simply requiring high bolometric luminosity and low \oiii EW, which can be plainly translated into high Eddington ratio and low inclination, respectively. The latter condition might introduce some bias against the detectability of mostly equatorial winds, but overall the opening angle is expected to be rather wide (e.g. Nardini et al. 2015). Notably, were its coordinates within the SDSS footprint, the same PDS\,456 would be retrieved. 

Accretion rate and orientation are known to be the major drivers of quasar diversity (e.g. Shen \& Ho 2014, and references therein). All the other distinctive attributes of \ton, such as strong \feii, narrow \hb, blueshifted \oiii and \civ, and prominent soft X-ray excess, naturally follow from their mutual correlations within the Eigevenctor 1 formalism. Most of these features are observed in efficiently accreting AGN at both low and high redshift, with some variance possibly due to the line-of-sight inclination (cf. Jin et al. 2017; Vietri et al. 2018). In the Eigenvector 1 space, \ton belongs to the `blue outliers' sub-class of the extreme population A sources (Zamanov et al. 2002). In a future work, we will revisit its multiwavelength continuum and line properties to investigate the role of ultra-fast X-ray winds in these exceptional objects, which might shed new light on the physics behind the observed Eigenvector 1 correlations. 

\section*{Acknowledgments}
We thank the referee, M. Giustini, for helpful comments. This work is based on observations obtained with the ESA science mission \xmm, with instruments and contributions directly funded by ESA member states and NASA. EN acknowledges funding from the European Union's Horizon 2020 research and innovation programme under the Marie Sk\l{}odowska-Curie grant agreement no. 664931. EL is supported by an EU COFUND/Durham Junior Research Fellowship under grant agreement no. 609412. EL also acknowledges funding from the EU Horizon 2020 AHEAD project under grant agreement no. 654215, and thanks C. Vignali for guidance on X-ray data reduction. SB is supported by NASA through the \chandra award no. AR7-18013X issued by the \chandra X-ray Observatory Center, which is operated by the Smithsonian Astrophysical Observatory for and on behalf of NASA under contract NAS8-03060. 



\label{lastpage}

\end{document}